\documentclass[10pt,conference]{IEEEtran}
\usepackage{epsfig,graphicx,subfigure,psfrag,amsmath,cases,bm}
\usepackage{latexsym,amssymb,amsmath,epsfig,subfigure,algorithm,mathtools}
\usepackage{algorithmic}
\usepackage{color}
\usepackage{url}
\usepackage{scrtime}
\usepackage{stfloats}
\usepackage[left=0.62in,top=0.7in,bottom=0.65in,right=0.62in]{geometry}
\author{\textit{(Invited Paper)}\vspace*{2mm}\\\IEEEauthorblockN {Dongfang Xu, Xianghao Yu, and Robert Schober\\Friedrich-Alexander-University Erlangen-N\"urnberg, Germany\vspace*{-10mm}}

}

\newtheorem{T-Prob}{Transformed Problem}

\DeclareMathOperator{\maxo}{maximize}
\DeclareMathOperator{\mino}{minimize}

 \newcommand{\qed}{\hfill \ensuremath{\blacksquare}}

\title{Resource Allocation for Intelligent Reflecting Surface-Assisted Cognitive Radio Networks}

\begin{document}
\maketitle
\begin{abstract}
In this paper, we investigate resource allocation algorithm design for intelligent reflecting surface (IRS)-assisted multiuser
cognitive radio (CR) systems. In particular, an IRS is deployed to mitigate the interference caused by the secondary network to the primary users. The beamforming vectors at the base station (BS) and the phase shift matrix at the IRS are jointly optimized for maximization of the sum rate of the secondary system. The algorithm design is formulated as a non-convex optimization problem taking into account the maximum interference tolerance of the primary users. To tackle the resulting non-convex optimization problem, we propose an alternating optimization-based suboptimal algorithm exploiting
semidefinite relaxation, the penalty method, and successive convex approximation. Our simulation results show that the system sum rate is dramatically improved by our proposed scheme compared to two baseline schemes. Moreover, our results also illustrate the benefits of deploying IRSs in CR networks.
\end{abstract}
\section{Introduction}
Radio spectrum is naturally a limited resource in wireless communication systems. During the last couple of decades, most of the available spectrum has been licensed for providing high data-rate communication services. This has led to the spectrum scarcity problem for the fifth-generation and beyond wireless communication systems \cite{wong2017key}. On the other hand, measurements of the practical spectrum utilization have shown that a large amount of the licensed spectrum is highly underutilized \cite{spectrumreport}, \cite{datla2009spectrum}. As a remedy to improve spectral efficiency, communication systems employing cognitive radio (CR) technology have been emerged as a promising paradigm to provide communication services for unlicensed secondary systems without seriously degrading the system performance of the primary network \cite{datla2009spectrum}\nocite{4570202}--\cite{6522183}. For example, the authors of \cite{4570202} studied the joint transmit power allocation and receive beamforming design for minimization of the total transmit power in a CR network. In \cite{6522183}, the authors investigated the downlink (DL) beamforming algorithm design for minimization of the total transmit power while satisfying the quality-of-service (QoS) constraints of the secondary users (SUs) and limiting the interference leakage to the primary users (PUs) to be below a given interference threshold. However, since wireless channels are essentially random and largely uncontrollable, the designs proposed in \cite{4570202}, \cite{6522183} cannot effectively mitigate the interference to PUs in unfavorable radio frequency propagation environments. Therefore, a more effective interference management method is needed for reliable CR networks.
\par
Recently, intelligent reflecting surfaces (IRSs) have emerged as a promising solution for harnessing interference in wireless communication systems \cite{di2019smart}\nocite{yu2019miso,xu2019resource,pan2019intelligent,yu2019enabling}--\cite{zhang2019multiple}. In particular, comprising a set of passive phase shifters, an IRS is able to reflect the incident signals with desired phase shifts \cite{cui2014coding}. By smartly configuring the IRS, wireless channels can be proactively manipulated, which offers a high flexibility in resource allocation \cite{8910627}. Moreover, as desired, the reflected signals can be combined with non-reflected signals in a destructive or constructive manner to inhibit detrimental interference or enhance the desired signal power strength, which improves system performance without deploying additional costly and energy-consuming communication infrastructures. 
Noticing the high potential of IRSs many works have proposed the application of IRSs to boost the performance of communication systems \cite{yu2019miso}\nocite{xu2019resource,pan2019intelligent}--\cite{yu2019enabling}. 
Yet, the designs proposed in \cite{yu2019miso}\nocite{xu2019resource,pan2019intelligent}--\cite{yu2019enabling} either target single-user systems or handle the unit modulus constraint introduced by IRS by employing manifold optimization. As a result, these designs are not directly applicable to IRS-assisted multiuser CR networks, since IRS-assisted multiuser CR systems are more complex and the feasible sets of the corresponding optimization problems are not manifolds. In \cite{yuan2019intelligent}, the authors considered an IRS-aided CR system with variable magnitude IRS elements and proposed a joint beamforming and IRS design for maximization of the system sum rate. However, the proposed design in \cite{yuan2019intelligent} is not applicable to IRS-aided CR systems with unit magnitude IRS elements. To the best of the authors’ knowledge, the joint beamforming and IRS algorithm design for IRS-assisted multiuser CR networks with unit magnitude IRS elements is still an open issue. 
\par
Motivated by the above discussion, in this paper, we investigate resource allocation algorithm design for IRS-assisted CR communication systems. To this end, we maximize the sum rate of the secondary system by jointly optimizing the DL transmit beamformers at the BS and the phase shifts at the IRS. The formulated non-convex optimization problem is very challenging due to the unit-modulus constraint introduced by the IRS and the coupling between the optimization variables. Hence, the optimal solution for the considered problem is in general intractable. Therefore, in this paper, we propose an alternating optimization (AO)-based iterative suboptimal algorithm to handle the considered problem \cite{bezdek2002some}.
\par
\textit{Notations:} In this paper, boldface capital and lower case letters denote matrices and vectors, respectively. $\mathbb{N}$ denotes the set of nonnegative integers. $\mathbb{C}^N$ denotes the space of complex-valued vectors with length $N$. $\mathbb{R}^{N\times M}$ and $\mathbb{C}^{N\times M}$ denote the space of $N\times M$ real-valued and complex-valued matrices, respectively. $\Re\left \{ \cdot \right \}$ extracts the real value of a complex variable. $\mathbb{H}^{N}$ denotes the set of all $N$-dimensional complex Hermitian matrices. $\mathbf{I}_{N}$ is the $N\times N$ identity matrix. $|\cdot|$ and $||\cdot||_2$ denote the absolute value of a complex scalar and the $l_2$-norm of a vector, respectively. $\mathbf{x}^T$, $\mathbf{x}^*$, and $\mathbf{x}^H$ stand for the transpose, the conjugate, and the conjugate transpose of vector $\mathbf{x}$, respectively. $\mathbf{A}\succeq\mathbf{0}$ indicates that $\mathbf{A}$ is a positive semidefinite matrix. $\mathrm{Rank}(\mathbf{A})$, $\mathrm{Tr}(\mathbf{A})$, and $\left [ \mathbf{A} \right ]_{i,i}$ denote the rank, the trace, and the $(i,i)$-th entry of matrix $\mathbf{A}$, respectively. $x_i$ denotes the $i$-th element of vector $\mathbf{x}$. $\mathrm{Diag}(\mathbf{X})$ represents a diagonal matrix
whose diagonal elements are extracted from the main diagonal of matrix $\mathbf{X}$; $\mathrm{diag}(\mathbf{x})$ denotes an $N\times N$ diagonal matrix with diagonal elements $x_1, \cdots, x_N$. $\mathcal{E}\left \{ \cdot \right \}$ denotes statistical expectation. $\sim$ and $\overset{\Delta }{=}$ mean ``distributed as'' and ``defined as'', respectively. The distribution of a circularly symmetric complex Gaussian random variable with mean $\mu$ and variance $\sigma^2$ is denoted by $\mathcal{CN}(\mu ,\sigma^2)$. The gradient vector of function $f(\mathbf{x})$ with respect to $\mathbf{x}$ is denoted by $\nabla_{\mathbf{x}} f(\mathbf{x})$. $\mathbf{x}^{\dagger}$ denotes the optimal value of optimization variable $\mathbf{x}$.
\section{System Model}
\begin{figure}[t]
\vspace*{0mm}
\centering
\includegraphics[width=3in]{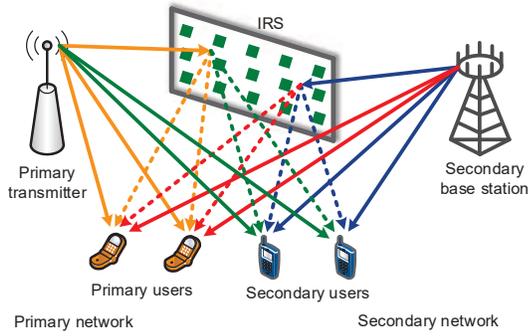} \vspace*{-2mm}
\caption{An intelligent reflecting surface (IRS)-assisted cognitive radio system.}
\label{system_model}\vspace*{-4mm}
\end{figure}
The considered IRS-assisted CR communication system comprises a primary license-holding network and a secondary unlicensed network, cf. Figure \ref{system_model}. In particular, the primary network comprises one primary transmitter and $I$ PUs while the secondary network is composed of one secondary BS and $K$ SUs. The secondary BS is equipped with $N_{\mathrm{T}}>1$ antennas while the PUs and SUs are single-antenna devices. Due to the spectrum sharing, the QoS of the PUs is impaired by the interference leakage from the secondary network. To effectively suppress the interference and boost the system performance of the secondary network, a passive IRS is deployed in the considered system. In particular, the IRS comprises $M$ phase-shifting elements, indexed by $\mathcal{M}\overset{\Delta }{=}\left \{1,\cdots ,M \right \}$, and is programmable and reconfigurable by an IRS controller. 
For notational simplicity, we define sets $\mathcal{I}=\left \{1,\cdots ,I \right \}$ and $\mathcal{K}=\left \{1,\cdots ,K \right \}$ to collect the indices of the corresponding users. Furthermore, we assume that perfect channel state information (CSI) of the whole system is available at the secondary BS for resource allocation design\footnote{In practice, the secondary BS may not be able to obtain perfect CSI of the whole CR system. Hence, the results in this paper serve as a theoretical system performance benchmark.}.
\par
The received signals at PU $i$ and SU $k$ are given by
\begin{eqnarray}
y_i^{\mathrm{P}}\hspace*{-2mm}&\hspace*{-2mm}=&\hspace*{-3mm}s_i^{\mathrm{P}}+\underbrace{\underset{k\in\mathcal{K} }{\sum }\hspace*{1mm}\mathbf{l}^H_{\mathrm{D},i}\mathbf{w}_kd_k+\underset{k\in\mathcal{K} }{\sum }\hspace*{1mm}\mathbf{l}^H_{\mathrm{R},i}\bm{\Psi} \mathbf{F}\mathbf{w}_kd_k}_{\text{ interference leakage from secondary network}}+n_i^{\mathrm{P}},\\
y_k^{\mathrm{S}}\hspace*{-2mm}&\hspace*{-2mm}=&\hspace*{-3mm}\underbrace{\mathbf{g}^H_{\mathrm{D},k}\mathbf{w}_kd_k+\mathbf{g}^H_{\mathrm{R},k}\bm{\Psi}\mathbf{F}\mathbf{w}_kd_k}_{\text{desired signal}}\notag\\&+&\underbrace{\underset{r\in\mathcal{K}\setminus \left \{ k \right \} }{\sum }\mathbf{g}^H_{\mathrm{D},k}\mathbf{w}_rd_r+\underset{r\in\mathcal{K}\setminus \left \{ k \right \} }{\sum }\mathbf{g}^H_{\mathrm{R},k}\bm{\Psi}\mathbf{F}\mathbf{w}_rd_r}_{\text{multiuser interference}}+n_k,
\end{eqnarray}
respectively, where $s_i^{\mathrm{P}}$ denotes the received signal originating from the primary transmitter. Moreover, $d_k\in\mathbb{C}$ denotes the information symbol for SU $k$ and $\mathbf{w}_k\in \mathbb{C}^{\mathit{N}_{\mathrm{T}}}$ is the corresponding beamformer. We assume $\mathcal{E}\{\left |d_k \right|^2\}=1$, $\forall\mathit{k} \in \mathcal{K}$, without loss of generality. The channel vector between PU $i$ and the secondary BS and the channel vector between PU $i$ and the IRS are denoted by $\mathbf{l}_{\mathrm{D},i}\in \mathbb{C}^{\mathit{N}_{\mathrm{T}}}$ and $\mathbf{l}_{\mathrm{R},i}\in \mathbb{C}^M$, respectively. Matrix $\mathbf{\Psi}=\mathrm{diag}\left ( e^{j\psi_1}, \cdots, e^{j\psi_M}  \right )$ represents the phase shift matrix of the IRS \cite{8811733}, \cite{wu2019beamforming} with $\psi_m$, $\forall m \in \mathcal{M}$, denoting the phase shift of the $m$-th reflector of the IRS. The channel between the secondary BS and the IRS is denoted by matrix $\mathbf{F}\in\mathbb{C}^{\mathit{M}\times\mathit{N}_{\mathrm{T}}}$. $\mathbf{g}_{\mathrm{D},k}\in \mathbb{C}^{\mathit{N}_{\mathrm{T}}}$ and $\mathbf{g}_{\mathrm{R},k}\in \mathbb{C}^M$ denote the channel vector between SU $k$ and the secondary BS and the channel vector between SU $k$ and the IRS, respectively. $n_k\sim\mathcal{CN}(0,\sigma_{\mathrm{n}_k}^2)$ is the equivalent noise at SU $k$, which captures the joint effect of the received interference from the primary network and thermal noise. $n_i^{\mathrm{P}}\sim\mathcal{CN}(0,\sigma_{\mathrm{n}_i}^2)$ represents the additive white Gaussian noise at PU $i$.

\section{Optimization Problem Formulation}
In this section, we formulate the resource allocation optimization problem for the considered system, after defining the adopted system performance metric.
\par
The achievable rate (bits/s/Hz) of SU $k$ is given by $R_k=\mathrm{log}_2(1+\Gamma_k)$, where $\Gamma_k$
is the received signal-to-noise-plus-interference ratio (SINR) of SU $k$ and is given as follows
\begin{equation}
\label{downlinkR}
\Gamma_k=\frac{\left | \mathbf{g}_{\mathrm{D},k}^H\mathbf{w}_k+\mathbf{g}_{\mathrm{R},k}^H \mathbf{\Psi}\mathbf{F}\mathbf{w}_k\right |^2}{\underset{r\in\mathcal{K}\setminus \left \{ k \right \} }{\sum} \left |\mathbf{g}_{\mathrm{D},k}^H\mathbf{w}_r+\mathbf{g}_{\mathrm{R},k}^H \mathbf{\Psi}\mathbf{F}\mathbf{w}_r\right |^2 +\sigma^2_{\mathrm{n}_k}}.
\end{equation}
\par
In this paper, we aim to maximize the system sum rate of the secondary network while limiting the interference leakage to the PUs below a threshold by optimizing $\mathbf{w}_k$ and $\mathbf{\Psi}$. The corresponding optimization problem is formulated as
\vspace*{-1mm}
\begin{eqnarray}
\label{prob1}
&&\hspace*{2mm}\underset{\mathbf{w}_k,\mathbf{\Psi}}{\maxo} \,\, \,\, F\big(\mathbf{w}_k,\mathbf{\Psi}\big)\overset{\Delta }{=}\underset{ k\in\mathcal{K}}{\sum}\mathrm{log}_2(1+\Gamma_k)\\
\mbox{s.t.}\hspace*{-6mm}
&&\mbox{C1:~}\underset{k\in\mathcal{K}}{\sum }\left \| \mathbf{w}_k \right \|^2\hspace*{-0.5mm}\leq\hspace*{-0.5mm} P^{\mathrm{max}},\mbox{C2:~}\mathbf{\Psi}\hspace*{-0.5mm}=\hspace*{-0.5mm}\mathrm{diag}\left ( e^{j\psi_1}\hspace*{-0.5mm}, \cdots\hspace*{-0.5mm}, e^{j\psi_M}  \right ),\notag\\
&&\mbox{C3:~}\underset{k\in \mathcal{K} }{\sum }\left | \mathbf{l}_{\mathrm{D},i}^H\mathbf{w}_k +\mathbf{l}_{\mathrm{R},i}^H\mathbf{\Psi }\mathbf{F}\mathbf{w}_k \right |^2\leq
p_{\mathrm{tol}_i},~\forall i\notag.
\end{eqnarray}
Here, constant $P^{\mathrm{max}}$ in constraint C1 represents the maximum transmit power allowance of the secondary BS. Constraint C2 guarantees that the phase shift matrix is a diagonal matrix with $M$ unit modulus components. C3 is the interference leakage constraint. In particular, the secondary network is required to control the interference leakage such that the maximum received interference power at PU $i$ does not exceed a given interference tolerance $p_{\mathrm{tol}_i}$.
\par
Due to the coupling between $\mathbf{w}_k$ and $\mathbf{\Psi}$, the fractional objective function, and the unit-magnitude constraint C2, \eqref{prob1} is a highly non-convex optimization problem and the optimal solution is in general intractable. Therefore, we propose an AO-based iterative suboptimal algorithm for finding a stationary point of \eqref{prob1}.

\section{Solution of the Problem}
In this section, an AO-based algorithm is developed to solve \eqref{prob1} in an alternating manner. In particular, by employing SCA and SDR, we first obtain the transmit beamforming vector $\mathbf{w}_k$ for a given $\mathbf{\Psi}$. Then, given $\mathbf{w}_k$, we solve for $\mathbf{\Psi}$ by applying a penalty-based method and SCA. 
\subsection{Optimizing $\mathbf{w}_k$ for Given $\mathbf{\Psi}$}
To facilitate resource allocation algorithm design, we first define $\mathbf{W}_k=\mathbf{w}_k\mathbf{w}_k^H$. Then, for given $\bm{\Psi}$, we first rewrite the terms $\left | \mathbf{g}_{\mathrm{D},k}^H\mathbf{w}_r+\mathbf{g}_{\mathrm{R},k}^H \mathbf{\Psi}\mathbf{F}\mathbf{w}_r\right |^2$  and $\left | \mathbf{l}_{\mathrm{D},i}^H\mathbf{w}_k +\mathbf{l}_{\mathrm{R},i}^H\mathbf{\Psi }\mathbf{F}\mathbf{w}_k \right |^2$ as follows,
\vspace*{-1mm}
\begin{eqnarray}
&&\hspace*{-6mm}\left | \mathbf{g}_{\mathrm{D},k}^H\mathbf{w}_k+\mathbf{g}_{\mathrm{R},k}^H \mathbf{\Psi}\mathbf{F}\mathbf{w}_k\right |^2=\left |\widetilde{\mathbf{g}}_k^H\mathbf{w}_k \right |^2=\mathrm{Tr}(\widetilde{\mathbf{g}}_k\widetilde{\mathbf{g}}_k^H\mathbf{W}_k),\\
&&\hspace*{-6mm}\left | \mathbf{l}_{\mathrm{D},i}^H\mathbf{w}_k +\mathbf{l}_{\mathrm{R},i}^H\mathbf{\Psi }\mathbf{F}\mathbf{w}_k \right |^2=\left |\widetilde{\mathbf{l}}_i^H\mathbf{w}_k \right |^2=\mathrm{Tr}(\widetilde{\mathbf{l}}_i\widetilde{\mathbf{l}}_i^H\mathbf{W}_k),
\end{eqnarray}
where $\widetilde{\mathbf{g}}_k$, $\widetilde{\mathbf{l}}_i\in \mathbb{C}^{N_{\mathrm{T}}\times 1}$ are defined as $\widetilde{\mathbf{g}}_k=\mathbf{g}_{\mathrm{D},k}+ \mathbf{F}^H\mathbf{\Psi}^H\mathbf{g}_{\mathrm{R},k}$ and $\widetilde{\mathbf{l}}_i=\mathbf{l}_{\mathrm{D},i}+\mathbf{F}^H\mathbf{\Psi }^H\mathbf{l}_{\mathrm{R},i}$. Then, the received SINR of SU $k$ is given by
\begin{equation}
\Gamma_k=\frac{\mathrm{Tr}(\widetilde{\mathbf{g}}_k\widetilde{\mathbf{g}}_k^H\mathbf{W}_k)}{\underset{r\in\mathcal{K}\setminus \left \{ k \right \} }{\sum} \mathrm{Tr}(\widetilde{\mathbf{g}}_k\widetilde{\mathbf{g}}_k^H\mathbf{W}_r)+\sigma^2_{\mathrm{n}_k}}.
\end{equation}
Moreover, constraint C3 can be rewritten equivalently as:
\begin{equation}
    \widehat{\mbox{C3}}\mbox{:~}\underset{k\in \mathcal{K} }{\sum }\mathrm{Tr}(\widetilde{\mathbf{l}}_i\widetilde{\mathbf{l}}_i^H\mathbf{W}_k)\leq p_{\mathrm{tol}_i},~\forall i.
\end{equation}
\par
Given $\mathbf{\Psi}$, the optimization problem design of the beamforming policy $\mathbf{W}_k$ is given as follows:
\begin{eqnarray}
\label{prob2}
&&\hspace*{-2mm}\underset{\mathbf{W}_k\in\mathbb{H}^{\mathit{N}_{\mathrm{T}}}}{\maxo} \,\, \,\, \underset{ k\in\mathcal{K}}{\sum}\mathrm{log}_2(1+\Gamma_k)\\
\mbox{s.t.}\hspace*{-2mm}
&&\mbox{C1:~}\underset{k\in\mathcal{K}}{\sum }\mathrm{Tr}(\mathbf{W}_k)\leq P^{\mathrm{max}},\hspace*{1mm}\widehat{\mbox{C3}},\notag\\
&&\mbox{C4:~}\mathbf{W}_k\succeq\mathbf{0},~\forall k,\hspace*{4mm}\mbox{C5:~}\mathrm{Rank}(\mathbf{W}_k)\leq1,~\forall k,\notag
\end{eqnarray}
where constraints C4, C5, and $\mathbf{W}_k\in\mathbb{H}^{\mathit{N}_{\mathrm{T}}}$ are imposed to guarantee that $\mathbf{W}_k=\mathbf{w}_k\mathbf{w}_k^H$ holds after optimization. Due to the objective function and the rank constraint C5, \eqref{prob2} is a non-convex problem. Next, we handle the optimization problem in \eqref{prob2} by applying SCA. To facilitate the application of SCA, we first define $f$ and $g$ which are given by
\begin{eqnarray}
&&f=-\underset{k\in\mathcal{K} }{\sum }\mathrm{log}_2\left (  \underset{r\in\mathcal{K} }{\sum }\mathrm{Tr}(\widetilde{\mathbf{g}}_k\widetilde{\mathbf{g}}_k^H\mathbf{W}_r)+\sigma^2_{\mathrm{n}_k}\right ),\\
&&g=-\underset{k\in\mathcal{K} }{\sum }\mathrm{log}_2\left (\underset{r\in\mathcal{K}\setminus \left \{ k \right \} }{\sum}\mathrm{Tr}(\widetilde{\mathbf{g}}_k\widetilde{\mathbf{g}}_k^H\mathbf{W}_r)+\sigma^2_{\mathrm{n}_k}\right).
\end{eqnarray}
We note that the objective function of \eqref{prob2} can be written as $g-f$.
Then, for any feasible point $\mathbf{W}^{(j)}_k$, we construct the global underestimator of $g(\mathbf{W}_k)$ which is given by
\begin{eqnarray}
\label{G1s}
g(\mathbf{W}_k)&&\hspace*{-6mm}\geq g(\mathbf{W}^{(j)}_k)+\underset{k\in\mathcal{K}}{\sum}\mathrm{Tr}\Big(\big(\nabla_{\mathbf{W}_k}g(\mathbf{W}^{(j)}_k)\big)^H(\mathbf{W}_k-\mathbf{W}_k^{(j)})\Big)\notag\\
&&\hspace*{-6mm}\overset{\Delta }{=}\widehat{g}(\mathbf{W}_k,\mathbf{W}_k^{(j)}),
\end{eqnarray}
where $\nabla_{\mathbf{W}_k}g(\mathbf{W}_k)$ is given by
\begin{equation}
\nabla_{\mathbf{W}_k}g(\mathbf{W}_k)=-\frac{1}{\mathrm{ln2}}\underset{t\in\mathcal{K}\setminus \left \{ k \right \} }{\sum}\frac{\mathbf{g}_k\widetilde{\mathbf{g}}_k^H}{\underset{r\in\mathcal{K}\setminus \left \{ t \right \} }{\sum}\mathrm{Tr}(\widetilde{\mathbf{g}}_k\widetilde{\mathbf{g}}_k^H\mathbf{W}_r)+\sigma^2_{\mathrm{n}_k}}.
\end{equation}
\par
Then, for a given feasible point $\mathbf{W}^{(j)}_k$, we solve the following problem:
\begin{eqnarray}
\label{prob4}
&&\hspace*{-22mm}\underset{\mathbf{W}_k\in\mathbb{H}^{\mathit{N}_{\mathrm{T}}}}{\mino} \,\, \,\, f-\widehat{g}(\mathbf{W}_k,\mathbf{W}_k^{(j)})\\
\mbox{s.t.}\hspace*{-2mm}
&&\mbox{C1},\widehat{\mbox{C3}},\mbox{C4},\mbox{C5},\notag
\end{eqnarray}
The only non-convexity in \eqref{prob4} results from rank constraint C5. By adopting SDR, we remove constraint C5 and the relaxed version of problem \eqref{prob4} can be optimally solved by applying convex solver CVX \cite{grant2008cvx}. Next, we reveal the tightness of SDR by presenting the following theorem.
\par
\textit{Theorem 1:~}If $P^{\mathrm{max}}>0$, the optimal beamforming matrix $\mathbf{W}_k$ always satisfies $\mathrm{Rank}(\mathbf{W}_k)\leq 1$.
\par
\textit{Proof:~}Problem (\ref{prob4}) is similar to \cite [Problem (15)]{xu2019resource} and the proof of Theorem 1 closely follows \cite [Appendix]{xu2019resource}. Hence, we omit the details of the proof due to space constraints. \qed
\par
The solution of \eqref{prob4} for given $\mathbf{W}^{(j)}_k$ will be integrated into the overall AO algorithm in Section IV-C.
\subsection{Optimizing $\mathbf{\Psi}$ for Given $\mathbf{w}_k$}
For given $\mathbf{w}_k$, the optimization problem for IRS design is given by
\vspace*{-1mm}
\begin{eqnarray}
\label{prob7}
&&\hspace*{-16mm}\underset{\mathbf{\Psi}}{\maxo} \,\, \,\, \underset{ k\in\mathcal{K}}{\sum}\mathrm{log}_2(1+\Gamma_k)\\
\mbox{s.t.}\hspace*{-2mm}
&&\mbox{C2:~}
\mathbf{\Psi}=\mathrm{diag}\left ( e^{j\psi_1}, \cdots, e^{j\psi_M}  \right ),~\mbox{C3}\notag.
\end{eqnarray}
We note that both the objective function and constraint C2 are non-convex functions which makes IRS design very challenging. Next, we first tackle the non-convex objective function in \eqref{prob7}. In particular, we rewrite the quadratic term $\left | \mathbf{g}_{\mathrm{D},k}^H\mathbf{w}_r+\mathbf{g}_{\mathrm{R},k}^H \mathbf{\Psi}\mathbf{F}\mathbf{w}_r\right |^2$ in \eqref{downlinkR} as follows:
\begin{eqnarray}
&&\left | \mathbf{g}_{\mathrm{D},k}^H\mathbf{w}_r+\mathbf{g}_{\mathrm{R},k}^H \mathbf{\Psi}\mathbf{F}\mathbf{w}_r\right |^2\notag
\\&&\hspace*{-6mm}=\mathrm{Tr}\Big(\hspace*{-1mm}\begin{bmatrix}
\bm{\theta}^H \hspace*{-1mm}&\hspace*{-1mm} \rho^*
\end{bmatrix}\hspace*{-1mm}\begin{bmatrix}
\mathrm{diag}(\mathbf{g}_{\mathrm{R},k}^H)\mathbf{F}\\ \mathbf{g}^H_{\mathrm{D},k}
\end{bmatrix}\hspace*{-1mm}\mathbf{W}_r\hspace*{-1mm}\begin{bmatrix}
\mathbf{F}^H\mathrm{diag}(\mathbf{g}_{\mathrm{R},k})\hspace*{-1mm}&\hspace*{-1mm} \mathbf{g}_{\mathrm{D},k}
\end{bmatrix}
\hspace*{-1mm}\begin{bmatrix}
\bm{\theta}\\ \rho
\end{bmatrix}\Big)\notag
\\&&\hspace*{-6mm}=\mathrm{Tr}(\widetilde{\bm{\theta}}^H\mathbf{G}_{k}\mathbf{W}_r\mathbf{G}_{k}^H\widetilde{\bm{\theta}} )=\mathrm{Tr}(\bm{\Theta}\mathbf{G}_{k}\mathbf{W}_r\mathbf{G}_{k}^H),
\end{eqnarray}
where optimization variables $\bm{\theta}\in\mathbb{C}^{M\times 1}$, $\widetilde{\bm{\theta}}\in\mathbb{C}^{(M+1)\times 1}$, and $\bm{\Theta}\in\mathbb{C}^{(M+1)\times(M+1)}$ are defined as $\bm{\theta}=[e^{j\psi_1}, \cdots, e^{j\psi_M}]^H$, $\widetilde{\bm{\theta}}=[\bm{\theta}^T~ \rho]^T$, and $\bm{\Theta}=\widetilde{\bm{\theta}}\widetilde{\bm{\theta}}^H$, respectively. Moreover, $\rho\in\mathbb{C}$ is a dummy variable with $\left | \rho \right |=1$. Besides, $\mathbf{G}_{k}\in\mathbb{C}^{(M+1)\times N_{\mathrm{T}}}$ is defined as  $\mathbf{G}_{k}=\big[\big(\mathrm{diag}(\mathbf{g}_{\mathrm{R},k}^H)\mathbf{F}\big)^T~~\mathbf{g}_{\mathrm{D},k}^*\big]^T$. Then, the received SINR of SU $k$ can be equivalently rewritten as follows:
\begin{equation}
\label{downlinkR2}
\Gamma_k=\frac{\mathrm{Tr}(\bm{\Theta}\mathbf{G}_{k}\mathbf{W}_k\mathbf{G}_{k}^H)}{\underset{r\in\mathcal{K}\setminus \left \{ k \right \} }{\sum}\mathrm{Tr}(\bm{\Theta}\mathbf{G}_{k}\mathbf{W}_r\mathbf{G}_{k}^H)+\sigma^2_{\mathrm{n}_k}}.
\end{equation}
\par
Similarly, we rewrite constraint C3 equivalently as follows:
\begin{equation}
\widetilde{\mbox{C3}}\mbox{:}\underset{ k\in\mathcal{K}}{\sum}\mathrm{Tr}(\bm{\Theta}\mathbf{L}_{i}\mathbf{W}_k\mathbf{L}_{i}^H)\leq p_{\mathrm{tol}_i},~\forall i,
\end{equation}
where $\mathbf{L}_{i}\in\mathbb{C}^{(M+1)\times N_{\mathrm{T}}}$ is defined as $\mathbf{L}_{i}=\big[\big(\mathrm{diag}(\mathbf{l}_{\mathrm{R},i}^H)\mathbf{F}\big)^T~~\mathbf{l}_{\mathrm{D},i}^*\big]^T$.
\par
Next, we focus on tackling the non-convex objective function in \eqref{prob7}. To start with, we define $\widetilde{f}$ and $\widetilde{g}$ which are given by, respectively,
\begin{eqnarray}
\widetilde{f}&&\hspace*{-6mm}=-\underset{k\in\mathcal{K}}{\sum }\mathrm{log}_2\Big(\underset{r\in\mathcal{K}}{\sum}\mathrm{Tr}(\bm{\Theta}\mathbf{G}_{k}\mathbf{W}_r\mathbf{G}_{k}^H)+\sigma^2_{\mathrm{n}_k}\Big),\\
\widetilde{g}&&\hspace*{-6mm}=-\underset{k\in\mathcal{K}}{\sum }\mathrm{log}_2\Big(\underset{r\in\mathcal{K}\setminus \left \{ k \right \} }{\sum}\mathrm{Tr}(\bm{\Theta}\mathbf{G}_{k}\mathbf{W}_r\mathbf{G}_{k}^H)+\sigma^2_{\mathrm{n}_k}\Big).
\end{eqnarray}
\par
Therefore, for given $\mathbf{W}_k$, \eqref{prob7} is rewritten as the following non-convex problem
\vspace*{-1mm}
\begin{eqnarray}
\label{prob8}
&&\hspace*{-2mm}\underset{\mathbf{\Theta}\in\mathbb{H}^{M+1}}{\mino} \,\, \,\, \widetilde{f}-\widetilde{g}\\
\mbox{s.t.}\hspace*{-2mm}
&&\widetilde{\mbox{C2}}\mbox{:}~
\mathrm{Diag}(\mathbf{\Theta})=\mathbf{I}_{M+1},~\widetilde{\mbox{C3}}\notag\\
&&\mbox{C6:~}\mathbf{\Theta}\succeq\mathbf{0},~\mbox{C7:~}\mathrm{Rank}(\mathbf{\Theta})=1,\notag
\end{eqnarray}
where $\bm{\Theta}\succeq\mathbf{0}$ in constraint C6, $\mathrm{Rank}(\mathbf{\Theta})=1$ in constraint C7, and $\mathbf{\Theta}\in\mathbb{H}^{M+1}$ are imposed to ensure that $\bm{\Theta}=\widetilde{\bm{\theta}}\widetilde{\bm{\theta}}^H$ holds after optimization. We note that rank-one constraint C7 stemming from the unit-modulus constraint is an obstacle to solving problem \eqref{prob8}. In the literature, SDR is commonly adopted to tackle the rank-one constraint \cite{5447068}. Yet, one major issue when employing SDR is that the solution obtained for \eqref{prob8} may not be a unit rank matrix. Moreover, some approximation methods such as Gaussian randomization cannot guarantee the convergence of the overall AO algorithm \cite{bezdek2002some}. To tackle this obstacle, we first rewrite rank-one constraint C7 equivalently as the following constraint:
\begin{equation}
    \widetilde{\mbox{C7}}\mbox{:}~\left \|\bm{\Theta}\right \|_*-\left \| \mathbf{\Theta } \right \|_2\leq 0,
\end{equation}
where $\left \| \mathbf{\Theta } \right \|_2$ and $\left \|\bm{\Theta}\right \|_*$ denote the spectral norm and nuclear norm, respectively. In particular, $\left \| \mathbf{\Theta } \right \|_2$ is given by $\left \| \mathbf{\Theta } \right \|_2=\sigma_1(\mathbf{\Theta })$ where $\sigma_i(\mathbf{\Theta })$ denotes the $i$-th largest singular value of matrix $\mathbf{\Theta }$. We note that for any $\bm{\Theta}\in\mathbb{H}^{M+1}$ and $\bm{\Theta}\succeq\mathbf{0}$, we have $\left \|\bm{\Theta}\right \|_*=\underset{i}{\sum}~\sigma_i(\mathbf{\Theta })\geq\left \| \mathbf{\Theta } \right \|_2=\underset{i}{\mathrm{max}}~\sigma_i(\mathbf{\Theta })$ and the equality holds if and only if $\bm{\Theta}$ is a rank-one matrix. Yet, the resulting constraint $\widetilde{\mbox{C7}}$ is still non-convex. To tackle this obstacle, we adopt the penalty-based approach and recast \eqref{prob8} as follows:
\begin{eqnarray}
\label{prob9}
&&\hspace*{-26mm}\underset{\mathbf{\Theta}\in\mathbb{H}^{M+1}}{\mino} \,\, \,\, \widetilde{f}-\widetilde{g}+\chi\big(\left \|\bm{\Theta}\right \|_*-\left \| \mathbf{\Theta } \right \|_2\big)\\
\mbox{s.t.}\hspace*{4mm}
&&\widetilde{\mbox{C2}},\widetilde{\mbox{C3}},\mbox{C6},\notag
\end{eqnarray}
where $\chi>0$ is a constant which penalizes the objective function for any matrix $\bm{\Theta}$ whose rank exceeds one. The equivalence between problem \eqref{prob9} and problem \eqref{prob8} is revealed by the following theorem \cite{ben1997penalty}.
\par
\textit{Theorem 2:~} Denote the optimal solution of problem \eqref{prob9} for penalty factor $\chi_q$ by $\bm{\Theta}_q$. When $\chi_q$ is sufficiently large, i.e., $\chi_q\rightarrow \infty$,  then any limit point $\overline{\bm{\Theta}}$ of the sequence $\left \{ \bm{\Theta}_q \right \}$ is an optimal solution of problem \eqref{prob8}.
\par
\textit{Proof:~}Problem (\ref{prob9}) is similar to \cite [Problem (24)]{yu2019robust} and the proof of Theorem 2 closely follows \cite [Appendix C]{yu2019robust}. Hence, we omit the details of the proof due to space constraints. \qed
\par 
We note that the optimization problem in \eqref{prob9} is still a non-convex problem, due to the non-convex objective function. Yet, noticing that $\widetilde{f}$, $\widetilde{g}$, $\left \|\bm{\Theta}\right \|_*$, and $\left \| \mathbf{\Theta } \right \|_2$ are all convex functions, we can tackle the difference of convex (d.c.) programming problem in \eqref{prob9} by applying SCA. In particular, for any feasible point $\bm{\Theta}^{(j)}$, we construct a global underestimator of the differentiable convex function $\widetilde{g}$ which is given by
\begin{eqnarray}
\label{G1}
\hspace*{-4mm}\widetilde{g}(\bm{\Theta})&&\hspace*{-6mm}\geq \widetilde{g}(\bm{\Theta}^{(j)})\hspace*{-1mm}+\hspace*{-1mm}\mathrm{Tr}\Big(\big(\nabla_{\bm{\Theta}}\widetilde{g}(\bm{\Theta}^{(j)})\big)^H(\bm{\Theta}-\bm{\Theta}^{(j)})\Big)\\
&&\hspace*{-6mm}\overset{\Delta }{=}\overline{g}(\bm{\Theta},\bm{\Theta}^{(j)}),
\end{eqnarray}
where $\nabla_{\bm{\Theta}}\widetilde{g}$ is given by
\begin{equation}
   \nabla_{\bm{\Theta}}\widetilde{g}=-\frac{1}{\mathrm{ln2}}\underset{k\in\mathcal{K}}{\sum }\frac{\underset{r\in\mathcal{K}\setminus \left \{ k \right \} }{\sum}\mathbf{G}_{k}\mathbf{W}_r^H\mathbf{G}_{k}^H}{\underset{r\in\mathcal{K}\setminus \left \{ k \right \} }{\sum}\mathrm{Tr}(\bm{\Theta}\mathbf{G}_{k}\mathbf{W}_r\mathbf{G}_{k}^H)+\sigma^2_{\mathrm{n}_k}}, 
\end{equation}
and $\overline{g}(\bm{\Theta},\bm{\Theta}^{(j)})$ in \eqref{G1} is a global underestimator of $\widetilde{g}(\bm{\Theta})$. Similarly, for any feasible point $\bm{\Theta}^{(j)}$, we construct a global underestimator of $\left \| \mathbf{\Theta } \right \|_2$ which is given by \cite{yu2019robust}
\begin{eqnarray}
\label{Theta}
\hspace*{-4mm}\left \| \mathbf{\Theta } \right \|_2&&\hspace*{-6mm}\geq \left \| \mathbf{\Theta}^{(j)} \right \|_2\hspace*{-1mm}+\mathrm{Tr}\Big(\bm{\theta}^{(j)}_{\mathrm{max}}(\bm{\theta}^{(j)}_{\mathrm{max}})^H(\bm{\Theta}-\bm{\Theta}^{(j)})\Big)\hspace*{-1mm}\overset{\Delta }{=}\overline{\Theta}^{(j)},
\end{eqnarray}
where $\bm{\theta}^{(j)}_{\mathrm{max}}$ is the eigenvector corresponding to the maximum eigenvalue of matrix $\mathbf{\Theta }^{(j)}$.
\par
Therefore, for any given point $\mathbf{\Theta }^{(j)}$, an upper bound of \eqref{prob7} can be obtained by iteratively solving the following optimization problem:
\begin{eqnarray}
\label{prob10}
&&\hspace*{-26mm}\underset{\mathbf{\Theta}\in\mathbb{H}^{M+1}}{\mino} \,\, \,\, \widetilde{f}-\overline{g}(\bm{\Theta},\bm{\Theta}^{(j)})+\chi\big(\left \|\bm{\Theta}\right \|_*-\overline{\Theta}^{(j)}\big)\\
\mbox{s.t.}\hspace*{4mm}
&&\widetilde{\mbox{C2}},\widetilde{\mbox{C3}},\mbox{C6}.\notag
\end{eqnarray}
Now, the problem in \eqref{prob10} is a convex optimization problem which can be efficiently solved via CVX \cite{grant2008cvx}. 
\subsection{Overall AO Algorithm}
The overall AO based algorithm is summarized in \textbf{Algorithm 1}. Specifically, the minimum values of problems \eqref{prob4} and \eqref{prob10} serve as upper bounds for the optimal values of
problems \eqref{prob2} and \eqref{prob7}, respectively. In each iteration of Algorithm 1, the upper bound is tightened by solving \eqref{prob4} and \eqref{prob10} optimally. Moreover, we note that any limit point of the non-increasing sequence $\left \{\mathbf{w}_k^{(j)},\mathbf{\Psi}^{(j)}\right \}_{j\in\mathbb{N}}$ is a stationary point of problem \eqref{prob1} and the function value of sequence $\left \{\mathbf{w}_k^{(j)}, \mathbf{\Psi}^{(j)}\right \}_{j\in\mathbb{N}}$ is guaranteed to converge to a stationary value of the objective function of problem \eqref{prob1} \cite{bezdek2002some}.
\begin{algorithm}[t]
\caption{Alternating Optimization Based Algorithm}
\begin{algorithmic}[1]
\small
\STATE Set initial points $\mathbf{w}_k^{(1)}$ and $\mathbf{\Psi}^{(1)}$, iteration index $j=1$, and convergence tolerance $0\leq\varepsilon_{\mathrm{AO}}<1$
\REPEAT
\STATE Solve \eqref{prob4} for given $\mathbf{\Psi}=\mathbf{\Psi}^{(j)}$ and obtain $\mathbf{w}_k^{(j+1)}$
\STATE Update $\mathbf{w}_k=\mathbf{w}_k^{(j+1)}$
\STATE Solve \eqref{prob10} for given $\mathbf{w}_k$ and obtain $\mathbf{\Theta}^{(j+1)}$
\STATE Decompose $\mathbf{\Theta}^{(j+1)}=\widetilde{\bm{\theta}}^{(j+1)}(\widetilde{\bm{\theta}}^{(j+1)})^H$ and recover $\mathbf{\Psi}^{(j+1)}$
\STATE Set $j=j+1$
\UNTIL $\frac{\left | F\Big(\mathbf{w}_k^{(j)}, \mathbf{\Psi}^{(j)}\Big)-F\Big(\mathbf{w}_k^{(j-1)}, \mathbf{\Psi}^{(j-1)}\Big) \right |}{\left |F\Big(\mathbf{w}_k^{(j-1)}, \mathbf{\Psi}^{(j-1)}\Big)  \right |}\leq \varepsilon_{\mathrm{AO}}$, where $F(\cdot,\cdot)$ is defined in \eqref{prob1}
\STATE $\mathbf{w}_k^{\dagger}=\mathbf{w}_k^{(j)}$ and  $\mathbf{\Psi}^{\dagger}=\mathbf{\Psi}^{(j)}$
\end{algorithmic}
\end{algorithm}
\section{Simulation Results}
In this section, the system performance of the
proposed resource allocation scheme is evaluated via simulations. 
In particular, we assume that both the primary transmitter and the secondary BS are located at the center of a cell with radius 100 meter, respectively. The distance between the primary transmitter and the secondary BS is 180 meter. $I=2$ PUs and $K=2$ SUs are uniformly and randomly distributed in the corresponding cells, respectively. The IRS is deployed at the middle point of the line connecting the primary transmitter and the secondary FD BS\footnote{For the simulations, we assume that the IRS has identical distances to the primary transmitter and the secondary BS, respectively. In practice, the location of the IRS can be either chosen for convenience or optimized.}. The adopted parameter values are listed in Table \ref{tab:parameters}.
\begin{table}[t]\vspace*{0mm}\caption{System parameters adopted in simulations.}\vspace*{0mm}\label{tab:parameters}\footnotesize
\newcommand{\tabincell}[2]{\begin{tabular}{@{}#1@{}}#2\end{tabular}}
\centering
\begin{tabular}{|l|l|l|}\hline
    \hspace*{-1mm}$f_c$ & Carrier center frequency & $2.5$ GHz \\
\hline
    \hspace*{-1mm}$\alpha_p$ & Path loss exponent & $3.5$ \\
\hline
    \hspace*{-1mm}$P^{\mathrm{max}}$ & Max. transmit power of the secondary BS & $30$ dBm \\ 
\hline
    \hspace*{-1mm}$G_i$ & BS antenna gain & $10$ dBi \\
\hline
    \hspace*{-1mm}$p_{\mathrm{tol}_i}$ & Interference tolerance of PU $i$ & $-90$ dBm \\
\hline
    \hspace*{-1mm}$\sigma^2_{\mathrm{n}_k}$ & Equivalent noise power at SU $k$ & $-90$ dBm \\
\hline
    \hspace*{-1mm}$\varepsilon_{\mathrm{AO}}$ & AO error tolerance & $0.01$ \\
\hline
    \hspace*{-1mm}$\chi$ & Penalty factor & $10^{3}$\\
\hline
\end{tabular}
\vspace*{-4mm}
\end{table}
\par
We consider two baseline schemes for comparison. For baseline scheme 1, we adopt zero-forcing (ZF) beamforming at the secondary BS and generate the phases of the IRS in a random manner. In particular, we fix the direction of beamformer $\mathbf{w}_k$ for desired user $k$ such that it lies in the null spaces of all the other users’ channels. Then, we solve a problem similar to problem \eqref{prob1} where we also optimize the power allocated to SU $k$, i.e., $p_k\in\mathbb{R}$. For baseline scheme 2, we assume that an IRS is not deployed. Then, by applying SCA, we optimize the beamforming vector $\mathbf{w}_k$ for maximization of the system sum rate based on problem \eqref{prob2}.
\begin{figure}[t]\vspace*{0mm}
 \centering
\includegraphics[width=3.4in]{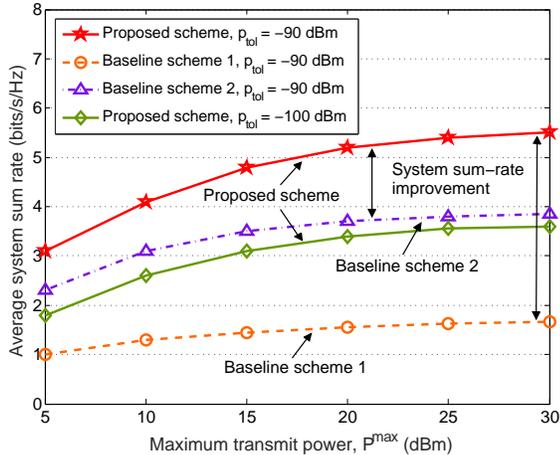}
\vspace*{-3mm}
\caption{Average system sum rate (bits/s/Hz) versus maximum transmit power (dBm) for different resource allocation schemes with $N_\mathrm{T}=8$, $M=8$, $I=2$, and $K=2$.}\vspace*{-5mm}\label{srpower}
\end{figure}
\par
In Figure \ref{srpower}, we study the average system sum rate versus the maximum transmit power at the secondary BS, $P^{\mathrm{max}}$, for different resource allocation schemes with $N_{\mathrm{T}}=8$, $M=8$, $I=2$, and $K=2$. As expected, with increasing $P^{\mathrm{max}}$, the system sum rates for the proposed scheme and the two baseline schemes increase monotonically. Moreover, we can observe that the proposed scheme outperforms the baseline schemes. In fact, compared to the baseline schemes, a considerable performance improvement of the proposed resource allocation scheme is enabled by the proposed joint optimization of $\mathbf{\Phi}$ and $\mathbf{w}_k$. As a result, the proposed scheme can simultaneously create a more favorable radio propagation environment and fully exploit the DoFs introduced by the multiplexing of multiple users. Moreover, by fully exploiting the available DoFs offered by the IRS, we can simultaneously enhance the desired signal while mitigating the interference. Besides, with decreasing interference tolerance $p_{\mathrm{tol}}$, the system sum rate of the proposed scheme decreases. This is due to the fact that to satisfy a lower interference tolerance, more DoFs have to be dedicated to suppressing the interference leakage to the PUs. As a result, fewer DoFs are available for enhancing the received power at the SUs which leads to a system performance degradation. The two baseline schemes achieve dramatically lower system sum rates. In particular, for baseline scheme 1, the secondary BS is unable to fully utilize the DoFs available for resource allocation since the ZF beamforming vector is fixed and the IRS phases are randomly generated. For baseline scheme 2, as there is no IRS, there are no DoFs available for establishing a favorable radio propagation environment.
\par
\begin{figure}[t]\vspace*{0mm}
 \centering
\includegraphics[width=3.4in]{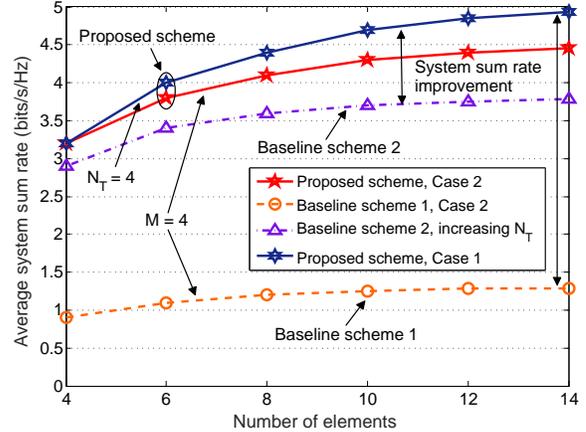}
\vspace*{-3mm}
\caption{Average system sum rate (bits/s/Hz) versus number of antenna/phase shifting elements for different resource allocation schemes with $K=2$, $I=2$, and $P^{\mathrm{max}}=20$ dBm.}\vspace*{-5mm}\label{srelement}
\end{figure}

In Figure \ref{srelement}, we investigate the average system sum rate versus the number of elements with $K=2$, $I=2$, and $P^{\mathrm{max}}=20$ dBm for different resource allocation schemes. In particular, to reveal the performance gain achieved by deploying IRSs, we compare the system performance of the proposed scheme for two cases: Case 1 with fixed $N_{\mathrm{T}}=4$ and increasing the number of IRS phase shifters $M$ and Case 2 with fixed $M=4$ and increasing the number of BS antennas $N_{\mathrm{T}}$. As can be seen from Figure \ref{srelement}, Case 1 (increasing the number of phase shifters at the IRS) produces a higher performance compared to Case 2 (increasing the number of antennas at the secondary BS). The reasons behind this are two-fold. On the one hand, the extra phase shifters provide higher flexibility in customizing the BS-IRS-user channels which improves the beamforming gain. On the other hand, more IRS elements can reflect more power of the signal transmitted by the BS which results in a power gain. Moreover, we can also observe that the average system sum rates for the proposed scheme and the two baseline schemes improve as $N_{\mathrm{T}}$ at the secondary BS increases. However, due to the channel hardening effect, the growth rate of the system sum rate gradually decreases for large values of $N_{\mathrm{T}}$.
\section{Conclusion}
In this paper, we proposed to integrate an IRS into a multiuser CR system to simultaneously enhance the system performance of the secondary network and effectively mitigate the interference to the PUs. In particular, we jointly optimize the transmit beamforming vectors and the phase shift matrix at the IRS for maximization of the system sum rate of the secondary network. Since the resulting optimization problem is highly non-convex, we developed an AO-based suboptimal algorithm to handle it in an alternating manner. Our simulation results show that the proposed scheme achieves a significant performance gain compared to two baseline schemes. Besides, our results also illustrate the benefits of deploying IRSs in CR networks.
\vspace*{-1mm}
\bibliographystyle{IEEEtran}
\bibliography{SPAWC_Refs}
\end{document}